# SNAC: An Unbiased Metric Evaluating Topology Recognize Ability of Network Alignment

**Hailong Li, Naiyue Chen**[*]
School of Computer and Information Technology, Beijing Jiaotong University, Beijing 100044, China
lihailong@bjtu.edu.cn, nychen@bjtu.edu.cn

**ABSTRACT**

Network alignment is a problem of finding the node mapping between similar networks. It links the data from separate sources and is widely studied in bioinformation and social network fields. The critical difference between network alignment and exact graph matching is that the network alignment considers node mapping in non-isomorphic graphs with error tolerance. Researchers usually utilize AC (accuracy) to measure the performance of network alignments which comparing each output element with the benchmark directly. However, this metric neglects that some nodes are naturally indistinguishable even in single graphs (e.g., nodes have the same neighbors) and no need to distinguish across graphs. Such neglect leads to the underestimation of models. We propose an unbiased metric for network alignment that takes indistinguishable nodes into consideration to address this problem. Our detailed experiments with different scales on both synthetic and real-world datasets demonstrate that the proposed metric correctly reflects the deviation of result mapping from benchmark mapping as standard metric AC does. Comparing with the AC, the proposed metric effectively blocks the effect of indistinguishable nodes and retains stability under increasing indistinguishable nodes.

---
[*] Corresponding author

# 1. INTRODUCTION

Network, or graph[1], can represent complex relationships of objects (e.g., article reference relationship, protein-protein interaction) because of its flexibility. However, flexibility sometimes exhibits an irregular side, leading to some problems related to graph challenges. As a classical problem of graph theory, graph isomorphism considers the bijection between the vertex sets of given graphs [1]: a bijection exists between node sets of given graphs if the graphs are isomorphic. Network alignment is a similar issue to finding bijection between graphs, whereas it allows non-isomorphism of input graphs. In addition to this, the mainstream models of network alignment tend to leverage statistics methods [2] or optimization method [3] to get approximate solution since the data network alignment models tackled often contains noise.

Because input graphs are often not isomorphic and do not exist perfect bijection between the nodesets, the output of network alignment cannot be perfect mapping and should not be evaluated by dichotomy (i.e., true or false) as the graph isomorphism situation adopted. In fact, virtually all models of network alignment use AC to evaluate their performances: for a pair of given benchmark mapping and the output mapping of model result, the way to compute AC is by counting consistent items between benchmark mapping and output mapping for nodes in given graphs, then divide the result by the number of total items [2][4].

This standard metric is intuitive and useful for evaluating models. Still, it neglects the specific case that some nodes in single graphs have identical topological structures and no need to distinguish. However, this identity is neither embodied in benchmark mapping nor output mapping and will introduce incorrectness to AC.

To this end, we propose SNAC (Symmetric Nodes Accuracy), a metric that takes identical nodes into consideration. SNAC treats nodes of the same topology indiscriminately when measuring the models.

This paper will focus on undirected networks without node attributes, while the proposed metric can be readily transplanted to directed graphs or attributed networks with some minor modifications. The main works and contributions of this paper are as follows.

- We define an equivalence relation called symmetry on the nodeset of a graph to judge if two nodes hold identical topology in the same graph. Based on node class sets of two graphs and node mapping across graphs, we define an alignment of nodes classes between graphs called ECM (equivalence classes mapping) to include symmetric nodes into benchmark mapping.
- We propose a novel metric SNAC for evaluating the deviation between node mapping and ECM across two networks. In terms of testing the ability of the models to recognize the network structures, the proposed metric yields unbiased measurements compared to AC. We give a theoretical explanation of the AC's deviation and give the deviation factor from AC to the model's actual ability to recognize network structure.
- We detailed describe the method utilizing an automorphism algorithm to find symmetric nodes. We introduce several measures leveraging the transitivity of equivalence relation into that method to improve computational efficiency. Then, we describe the method to alignment two classes of nodes to get ECM and give an adaptive solution under noise conditions.
- We conduct extensive experiments comparing metric SNAC with AC from two perspectives to show the exactitude and stability of SNAC. We also verify the above deviation factor of AC in experiments and give a guideline for choosing datasets when evaluating models under metric AC.

---

[1] This paper will use "graph" on theoretical context, and use "network" on practical context. However, these two terms also be used interchangeably in certain places.

The remainder of the paper is organized as follows. Section 2 gives an overview of network alignment, particularly explains the metrics evaluating the network alignment models. In Section 3, to set the stage for the subsequent definition, we first explain what symmetric nodes and equivalence classes mapping is, then build the definition of SNAC on given equivalence classes mapping and node mapping. Section 4 analyzes the root of metric AC's bias on models. Section 5 discusses the way of finding symmetric nodes and the way of calculating SNAC. Experiments and conclusions will be arranged in Section 6 and Section 7.

**Table 1. Symbols for the paper**

| Symbols | Definition |
| --- | --- |
| $G(V, E)$ | network with nodeset V and edgeset E |
| $G_1(V_1, E_1), G_2(V_2, E_2)$ | source network and target network |
| $\phi, \phi^*$ | result mapping and benchmark mapping on nodesets between $G_1$ and $G_2$ |
| $Aut(G)$ | the automorphism group of graph G |
| $\sigma$ | permutation of an automorphism |
| $\sim$ | symmetry relation on nodeset |
| $[n]$ | the equivalence class of node n |
| $V/\sim = \{[n] \mid n \in V\}$ | quotient set of V under $\sim$ |
| $\varphi(u), \varphi([u])$ | mapping of node u and mapping of class [u] |

## 2. RELATED WORKS

2.1 Network Alignment Overview

Network alignment is a task to find the node mapping of two networks, some literature [3][5] extend network alignment to multiple networks condition, here we consider only two networks case.

**Definition 1 Network Alignment** *Given two networks $G_1(V_1, E_1)$ and $G_2(V_2, E_2)$, design a network alignment method that aligns nodes and generates a node mapping $\phi: V_1 \rightarrow V_2$, let the errors between $\phi$ and true(benchmark) mapping $\phi^*$ as less as possible.*

The main symbols of this paper are listed in Table 1.

Network alignment is the critical step in applications that multiple networks are involved, including User Identity Linkage [6], Protein-protein interaction matching [3], Link prediction [7].

Models of network alignment can be classified in several ways, such as adjacent matrix factorization [3] based versus embedding based [2], utilizing only networks input [8] versus utilizing both networks and anchor links [9], single-level alignment versus multi-level alignment [10]. Recent works tend to employ machine learning techniques such as GAN [11] and GCN [12] to take advantage of AI and big data.

Generally, models produce a similarity matrix S of nodes across networks, and the alignments are carried out based on S. There are two ways to process this matrix. One [2][8] is to intuitively find the most similar counterparts in S for nodes of the source network, this way of alignment could map one node to multiple nodes in the target network. The other [13] is an iterative heuristic approach in which nodes were removed from the candidate lists once they have been aligned. By employing this approach, nodes only have one-to-one mapping. The second approach tends to produce a more accurate result comparing with the first one. However, it is more time-consuming. Some models [2][10] with a trivial precision gap between those two approaches adopt the former.

## 2.2 Network Alignment Evaluation

### 2.2.1 Evaluation Methods on Models

The evaluation methods on models vary with the input of models and vary with the way utilizing output alignment.

**Input-Based Evaluating Strategy**

Generally, models take a source network and a target network as input, and some models also take additional information called anchor links across networks. Based on the conditions of input networks and the existence of input anchor links, there are 3 cases.
1. Nodes in the source network are less than that of the target network, and each node in the source network has a counterpart in the target network [14]. In this case, metrics test if each node in the source network aligned correctly to its counterpart accords with benchmark mapping.
2. There is no quantitative restriction of networks nodes and some nodes in the source network may not have counterparts, such nodes are called gap nodes [3]. In this case, except for aligning nodes in $\phi$, models should mark out gap nodes.
3. In the case of models take anchor links [9][15][16], metrics will not evaluate anchor nodes in $\phi$.

**Output-Based Evaluating Metrics**

Following the previous work [17], we split mainstream metrics into two types based on the different output alignment utilizations.
1. Precision@$k$ (a.k.a. Success@$k$ [6]) indicates whether the correct match occurs in top-k candidates. Mention that Precision@1 is actually AC. Some works of literature call AC hard alignment and called Precision@$k$($k$>1) soft alignment [2].
2. Metric MAP [16] (a.k.a. Hit-Precision [4]) evaluates candidates from a ranking perspective, and it adds a trad-off to each correct match that occurs in top-$k$ candidates, given as:

$$MAP = \frac{1}{|\phi|} \sum_{a \epsilon \phi} \frac{1}{rank(a)} \qquad (1)$$

where *rank* is the position of the target node in the sorted list of candidates.

Our comparison will focus between SNAC and AC for conciseness, while the SNAC can be combined with Precision@$k$ or MAP to take the identical nodes into account.

### 2.2.2 Evaluation Method on Datasets

Since the performance of models is closely related to the consistency of given datasets, there exist metric reflecting the overlapping between networks.

*Interop* [18] reflects the topology consistency of two networks. The intuition of this metric is that two nodes should form an edge if their counterparts form an edge in another network. Models will hardly extract information from networks' structure if networks hold a low level of *Interop*. The definition of *Interop* is given below:

$$Interop(G_1, G_2) = \frac{|Correlations| * 2}{|E_1| + |E_2|} \qquad (2)$$

where *Correlations* are the overlapping edges of network $G_1$ and network $G_2$, $E_1$ and $E_2$ are the edgeset of $G_1$ and $G_2$, respectively.

Note that the by-product of computing SNAC also reflects the obstruction datasets brought to models. Different from *Interop* that considers two networks together, the by-product focuses on single networks. The details will be explained in Section4 and Section6.

## 2.3 Graph Automorphism

An automorphism [19] of a graph is a form of symmetry in which the graph is mapped onto itself while preserving edgeset unchanged. Formally, an automorphism of a graph $G$ $(V, E)$ is a permutation $\sigma$ of the nodes set $V$, such that the pair of nodes $(u, v)$ form an edge if and only if the pair $(\sigma(u), \sigma(v))$ also form an edge. There are many efficient algorithms to find graph automorphism [20][21][22].

We denote all automorphisms on graph $G$ as $Aut(G)$ in this paper, which is shown in Fig. 1. From the figure, we can see that the node pairs that transposed without altering the graph's topology cannot be differentiated, such nodes are the source of deviation in AC. We define these nodes as *symmetric nodes* and construct metric SNAC basing on *symmetric nodes* in this paper.

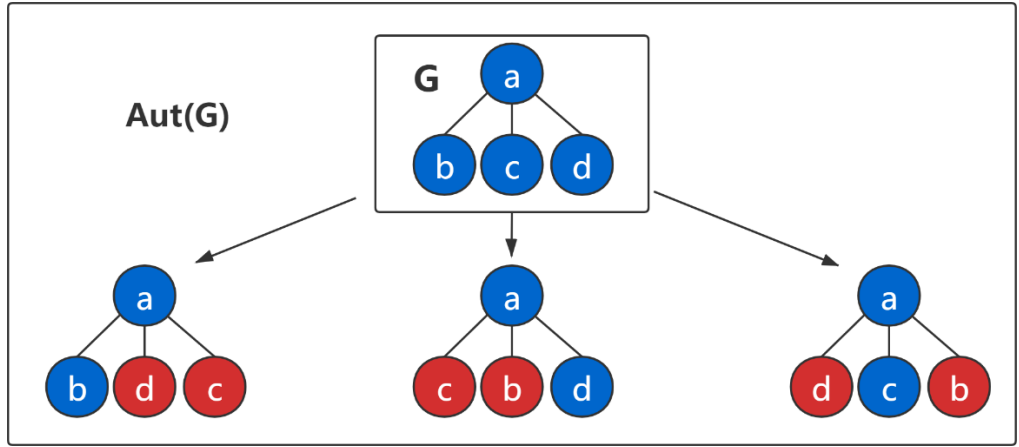

**Fig. 1. All automorphism of G, each of three derived graphs denotes a permutation σ of G, the transposed nodes are colored and hold identical topology.**

## 3. PROBLEM DEFINITION

### 3.1 Symmetric Nodes

Symmetric nodes are topologically identical to a graph. We define symmetric nodes as below.

**Definition 2 Symmetric nodes** *Let $Aut(G)$ denote the set of all automorphisms on a graph G, for every $\sigma \in Aut(G)$ and every $v \in V$, $(v, \sigma(v))$ is a pair of symmetric nodes. The relation of v and $\sigma(v)$ can be expressed as "v is symmetric to $\sigma(v)$" and vice versa.*

Since the graphs in $Aug(G)$ are isomorphic to each other and the graph isomorphism is an equivalence relation, from the above definition, we can see that the symmetry relation is also an equivalence relation which is reflexive, symmetric, and transitive. In this paper, we denote the symmetry relation of nodes $u$, $v$ by $u \sim v$.

The following Fig. 2. shows several cases of symmetric nodes. According to the definition of symmetric nodes, $g_1$'s nodeset $V = \{a, b, c, d\}$ have the equivalence relation $\{(a, a), (b, c), (c, d), (b, d), (d, b), (c, d), (d, c)\}$, The following sets are equivalence classes of this relation: $[a] = \{a\}$, $[b] = [c] = [d] = \{b, c, d\}$. The set of all equivalence classes (also called quotient set) for this relation is $\{\{a\}, \{b, c, d\}\}$, also denoted as $V/\sim$.

The equivalence relation and node quotient set of $g_2$ can be represented by $\{(a, a), (b, b), (c, d), (c, d)\}$ and $\{\{a\}, \{b\}, \{c, d\}\}$ in the same way of $g_1$. The $g_3$ has two groups of symmetric nodes and the equivalence relation and quotient set are $\{(a, a), (b, c), (c, b), (d, e), (e, d)\}$ and $\{\{a\}, \{b, c\}, \{d, e\}\}$, respectively.

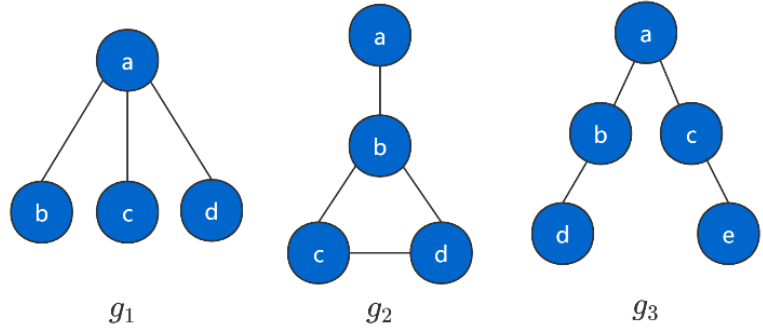

**Fig. 2. Examples of symmetric nodes**

3.2 ECM (Equivalence Class Mapping)

ECM is a mapping between two groups of equivalence classes. By converting one-to-one mapping $\phi^*$ to a many-to-many mapping ECM, the evaluating benchmark incorporates the symmetric nodes.

**Definition 3 ECM(Equivalence Class Mapping)** *For two given quotient sets $Q/\sim:=\{[x]|\ x \in V_1\}$, $P/\sim:=\{[y]|\ y \in V_2\}$ and node mapping $\phi^*: V_1 \rightarrow V_2$, equivalence class mapping aligns each class [x] to a class [y] based on $\phi^*$, denote the mapping as $\varphi([x])=[y]$.*

Fig. 3. depicts the ECM between networks S and T, mention that the benchmark mapping $\phi^*$ is *{(a, n), (b, o), (c, p), ..., (m, z)}*.

3.3 SNAC (Symmetric Nodes Accuracy)

SNAC tests the error between node mapping $\phi$ and ECM to measures the ability of models, here we formally define it.

**Definition 4 SNAC(Symmetric Nodes Accuracy)** *For each element $a=(u_1, u_2)$ of $\phi$, where $u_1 \in V_1$, $u_2 \in V_2$, the score of s(a) is:*

$$s(a) = \begin{cases} 1, & [u_2] = \varphi([u_1]), \\ 0, & other\ wise \end{cases}$$

And the SNAC:

$$\sum_{a \epsilon \phi} \frac{s(a)}{|\phi|} \qquad (3)$$

## 4. Metric Analysis

With the definition of symmetric node and SNAC, we can analyze the deviation of AC. To clearly describe, below we state in the context of one network and suppose that each node participates in evaluation.

Suppose the given model can correctly distinguish nodes from different topologies with probability $x$. Given a node $u$, we can get the equivalence class $[u]$ under the relation symmetry. Since the nodes in $[u]$ all have identical topology and is indistinguishable, the expectation of correctly identify $u$ in $[u]$ is $1/|[u]|$. Multiplying $x$ and $1/|[u]|$ to get $\frac{x}{|[u]|}$, it is the result precision of $u$ under the metric AC. The summation expectation score of class $[u]$ can be calculated by sum scores of all nodes in $[u]$:

$$\sum_{[u]} \frac{x}{|[u]|} = x$$

The total alignment score of all nodes $v \in V$ can be calculated by sum scores of all equivalence classes:

$$\sum_{V/\sim} x = x * |V/\sim|$$

Dividing the score by $|V|$ and get AC= $x * \frac{|V/\sim|}{|V|}$, we can see that the AC's deviation to $x$ is determined by $\frac{|V/\sim|}{|V|}$. Since $|V/\sim| \leqslant |V|$, AC tends to underestimate the models.

To SNAC, for a given node $u$, the expected probability of correctly identifying $u$ is remaining $x$ since SNAC can tackle the symmetric issue of nodes, we can get the score of all nodes by $\sum_V x = x * |V|$, and the $SNAC = x * \frac{|V|}{|V|} = x$.

To sum up, $SNAC = AC * \frac{|V|}{|V/\sim|}$ in statistics, and factor $\frac{|V|}{|V/\sim|}$ is the average length of all equivalence classes under the relation symmetry.

## 5. Method

The method of computing SNAC has three steps.

For two given networks, we first find the symmetric nodes in networks and organize them into the form of equivalence classes.

After separately organize the equivalence classes in the source network and target network, we can get ECM between that two groups of equivalence classes according to given nodes' mapping $\phi^*$. The ECM is independent of any alignment model and can be reutilized.

Finally, we can make utilization of ECM to evaluate the error of $\phi$ under the metric SNAC.

### 5.1 Finding Symmetric Nodes

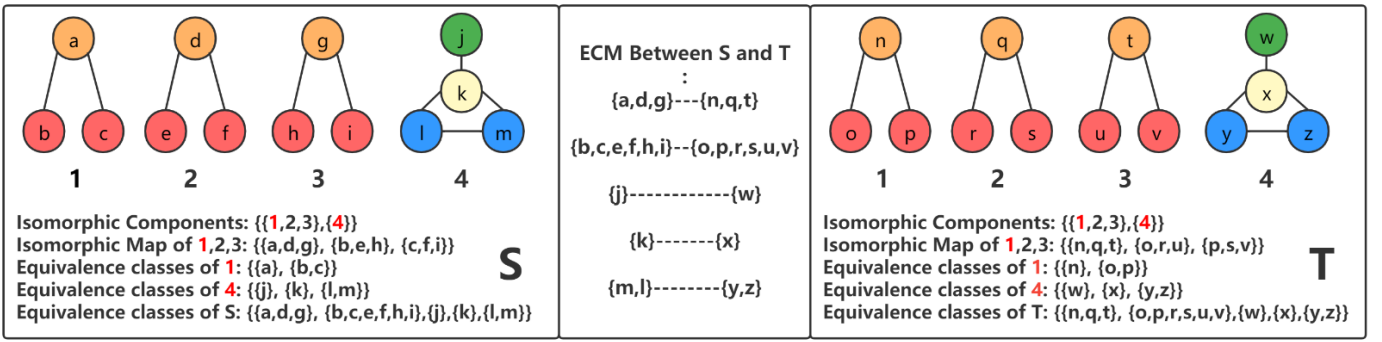

**Fig. 3. The ECM between S and T**

The automorphism problem involved in the process of finding symmetric nodes makes this process time-consuming. Therefore, we apply an improvement technique separating the given graph into several disconnected components, then leverage the transitivity of equivalence relation to merge the sub-results. This improvement strategy results in significant efficiency gains in practice, which is detailed in the appendix.

#### 5.1.1 Symmetric Nodes Across Components

For a given graph, it is sometimes not connected and has several components. Each pair of components can be isomorphic or not. If the former is true, then every node has at least one symmetric node in the other component. If the latter is true, there can be no symmetric nodes between two components, and we can find symmetric nodes in components independently. Because of the previous, the first step is to recognize isomorphic between components.

At the start, we employ deep-first search dividing graph into several disconnected components, then aggregate isomorphic components to the same groups. For each component pair in the same group, finding the isomorphic mapping of each node in them. Each mapping represents a pair of symmetric nodes. Aggregating isomorphic components and finding the mapping can be done simultaneously using *vf2* [23] algorithm.

In practice, testing isomorphism and finding symmetric nodes for every pair of components has quadratic complexity in *n*, where *n* is the number of components. The method described above can be optimized by utilizing the transitivity of the isomorphic relation.

We choose a landmark for each group. According to the transitive of isomorphic relation, to test whether a component belongs to a given group, we need only examine the isomorphism with the group's landmark. The symmetric nodes mapping of isomorphism sets can also be found in the processing of isomorphic testing because the symmetric relation of nodes is transitive too. By employing the optimizing strategy above, the complexity can approach $O(m^2)$, where $m(m<n)$ is the number of isomorphic sets.

### 5.1.2 Symmetric Nodes Within Components

For each component *g*, we employ *Bliss* [21] to produce an automorphism group of *g*, represented by several permutation matrixes. Symmetric nodes can be found in these matrixes since swapped rows and columns in matrixes denote symmetric relations.

Although *Bliss* adopts many pruning techniques to accelerate computation, the worst-case is also time-consuming. The algorithm's run time is unacceptable if we employ *Bliss* for every component. Fortunately, we can utilize the transitivity of the node's symmetric relation again.

**Algorithm 1: Finding Symmetric Nodes (G)**

```
1: ======== STEP1. Symmetric Nodes Across Components ======
2: Isomorphic_set=empty //set of isomorphic groups of graphs
3: Classes=empty   //equivalence classes list of all nodes in G, each element of
4:                 the list is a set of symmetric nodes.
5: for c in components of G do
6:     for group in Isomorphic_set do
7:         if is_isomorphic(group.landmark, c) then
8:             Classes|= isomorphic_mapping(group.landmark,c)
9:             // The operator "|=" merges two lists of sets, any two sets from
10:            // different list with intersections will be merged as one set by
11:            // the operator.
12:            group.add(c)
13:            Break
14:    if c is not isomorphic to any landmark of groups, then
15:        Isomorphic_set.add([c])
16:        Classes|= isomorphic_mapping(c,c)
17: ================ STEPS 2 and 3. ====================
18: ===== Symmetric Nodes Within Components and Combine ====
19: for group in Isomorphic set do
20:     Classes|= automorphic_mapping(landmark)
21: return Classes
```

### 5.1.3 Combine Symmetric Nodes

The nodes' symmetric mapping between landmarks and each component in the same isomorphic set has been obtained in section 5.1.1. So, we just need to compute automorphism groups for landmarks, then combine the mapping with symmetric mapping across components in the same set and get the ultimate equivalence classes.

The algorithm of finding symmetric nodes is shown in Algorithm 1. This improvement strategy divides the large-scale graph into several small-scale sub-graphs and conquers them separately. It is experimentally demonstrated in the appendix that it can significantly improve performance when the given graph contains many connected components.

Fig. 3. presents two examples of finding symmetric nodes, we choose the first one to explain: there are four components denoted as *1,2,3,4* in *S*. According to isomorphic relation, we divide them into two groups *{{1, 2, 3}, {4}}* and mark components 1,4 as landmarks. For each group, we first apply *vf2* to get the isomorphic mapping between components and landmark, then find symmetric nodes within the landmark, the equivalence classes of a group can be obtained by merging all symmetric nodes across components on transitivity. Ultimate equivalence classes of *S* can be obtained by directly concatenating equivalence classes of *{1, 2, 3}* and *{4}*. By doing this, we decompose the large-scale problem of recognizing symmetric nodes in *S* into several sub-problems and avoid directly computing the automorphism of sub-graphs 2,3.

**Algorithm 2: Computing ECM (Q, P, $\phi^*$)**

---

1: for source_class in Q do:

2:     assign 0 to each item in P to construct a map as mapping_candidate

3:     for v in source_class do

4:         incr mapping_candidate.value([$\phi^*(v)$])

5:     target_class=argmax(mapping_candidate)

6:     ECM.set_key_value(source_class, target_class)

7: return ECM

---

### 5.2 Compute ECM

The method of computing ECM between equivalence classes *Q* and *P* is according to benchmark mapping $\phi^*$ and is described below.

For every element $v_i \in [x], [x] \in Q$, finding its corresponding element $\phi^*(v_i)$ and $[\phi^*(v_i)]_j \in P$, counting the occurrence $n_j$ of $[\phi^*(v_i)]_j$ for each $v_i$. The class $[\phi^*(v)]_j$ that corresponding $max(n_j)$ is the mapping of class *[x]*.

The algorithm is shown in Algorithm2.

Notice that there is some inconsistency between equivalence classes of source graph and target graph in practice because of noise. We can incorporate this inconsistency by assigning weight to alignments between equivalence classes.

### 5.3 Compute SNAC

To evaluate the mapping $\phi$ based on equivalence class alignment, for a given element $(u_1, u_2) \in \phi$, where $u_1 \in V_1$, $u_2 \in V_2$, find the corresponding equivalence classes *[u₁], [u₂]*. The SNAC can be obtained by checking if $\varphi([u_1])$ is equal to *[u₂]* for each element of $\phi$.

The detail of SNAC's computation is described in section 3.3.

## 6. Experiments and Result

In this section, we answer four questions about our metrics. We conduct three tests based on three groups of datasets during each experiment, comparing SNAC and AC to illustrate the following points.

(Q1) Does SNAC accurately reflect the result alignment deviation from benchmark?

(Q2) Does SNAC block the effect of symmetric nodes in datasets and remain stable when evaluating models?

(Q3) Is there a property of datasets reflecting the effect symmetric nodes bringing to AC? How to get it?

(Q4) How do the mainstream network alignment models perform under metric SNAC?

6.1 Datasets

We use both real-world datasets and synthetic datasets in our experiments. Real-world datasets are used in basic ability test and models' performance test to show the basic ability and performance in the practice of SNAC and models. Synthetic datasets are used in stability test to show the feature of SNAC comparing to AC.

6.1.1 Real-world Dataset

Below we briefly describe the three real-world datasets we used in experiments.

(D1) **Collaboration Network** [24]: This dataset contains 18 772 nodes and 19 8110 edges. Collaboration network represents the collaboration of research works, each node in it is an author, and an edge between two authors denote that they are co-authors of some papers.

(D2) **PPI Network** [25]: This network contains 3890 nodes and 76584 edges. PPI network is often used in the biological field, where nodes and edges represent proteins and their interactions.

(D3) **Communication Network** [26]: This network represents email contacts where 1133 nodes and 5451 edges denote user and email between users, respectively.

6.1.2 Synthetic Dataset

To illustrate how standard metric AC is affected by symmetric nodes in the stability test, we use a random graph generator to produce networks.

The generator algorithm [27] constructs a network with two arguments *(n, p)* containing n nodes, every two nodes have an edge with probability p. Constructed networks have a favorable property that few symmetric nodes are involved in. We can leverage this to control the number of symmetric nodes by cloning the topology of nodes in constructed networks.

In the stability test, we generate three groups of datasets with the arguments (500, 0.04), (3000, 0.01), (5000, 0.005), each group contains six graphs with different symmetric node ratios from 0 to 0.25. For each dataset, we keep network topology the same and only shuffle the nodes' labels to generate the target network.

6.2 Network Alignment Models

To show that the proposed metric SNAC is generally applicable to models, we choose three well-known network alignment models using different strategies of algorithms.

(M1) **REGAL** [2]: REGAL utilizes the distribution of neighbor's degree as node feature and computes similarities between features as node embeddings. We can get the ultimate similarity matrix of nodes by computing the distance between the embeddings. It uses adjacency matrix factorization to reduce computation.

(M2) **IONE** [9]: IONE requires anchor links between the source network and target network to computes nodes embeddings. During iterations, each node takes embedding of the neighborhood as input to update its embedding. In our experiments, we randomly choose 10% from benchmark mapping $\phi^*$ as anchor links.

(M3) **IsoRank** [3]: IsoRank tackles network alignment as an integer quadratic problem with constraints and solves it by propagating nodes similarity between networks. IsoRank also takes prior information as IONE does. However, the information is a similarity matrix of nodes across networks and is not necessary. In our experiments, we construct that matrix from degree similarity between nodes.

The three models are different in several aspects, and the differences are shown in Table 2.

**Table 2. Differences of Models**

| Models | **REGAL** | **IONE** | **IsoRank** |
|---|---|---|---|
| Anchor Links | No | Yes | No |
| Confidence Propagation | No | Yes | Yes |
| Embedding Based | Yes | Yes | No |
| Support Nodes Attribute | Yes | No | No |
| Support Directed Graphs | No | Yes | No |

### 6.3 Basic Ability Test

The basic ability test aims to test if SNAC correctly reflects the error between $\phi$ and $\phi^*$ as AC does. This experiment is conducted directly on datasets without employing any network alignment models.

The three real-world datasets depicted in Section 6.1.1 are used in this experiment as three groups. We make a copy from the source network as the target network within each group, then randomly shuffle different ratios of node labels from 0.0 to 1.0 and construct result mapping. Since AC in this experiment does not involve topology issues, it will not be affected by symmetric nodes and can be a benchmark to test SNAC.

In Fig. 4., we see that both AC and SNAC have a good reflection of the discordance degree between the shuffled mapping and benchmark mapping, the gap from each bar of AC and SNAC to precision 1.0 is exactly the ratio of shuffled nodes. This simple experiment illustrates that SNAC has the essential ability as a metric to evaluate the error between mappings.

### 6.4 Stability Test

In the stability test, we will show the defect on metric AC and the advantage of SNAC. In addition, we give experimental evidence for the deduction in Section 4.

We use three network alignment models described in Section 6.2, and the models work on three groups of synthetic datasets as described in Section 6.1.2. Within each group of datasets, we keep all circumstances (network generator and parameters) the same except for increasing the ratio of symmetric nodes. By doing this, we can conveniently analyze the AC curve affected by symmetric nodes. In addition to AC and SNAC, we put a curve that multiplies AC by the average length of classes in Fig. 5. to make a comparison.

The three models in this experiment performed well because there is no noise between source and target networks, as shown in Fig. 5. However, the AC curve is visibly declining along with the increase of symmetric nodes, whereas SNAC keeps stable and keeps reflecting the models' ability to distinguish nodes without bias.

The additional curve goes around SNAC and fits better along with the scale and the density of networks, as we have argued in Section 4 from the mathematical view.

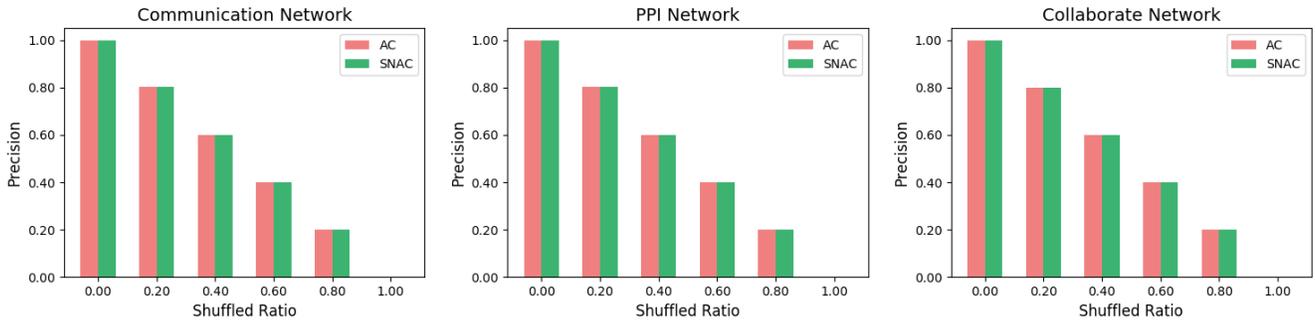

**Fig. 4. Basic Ability Test**

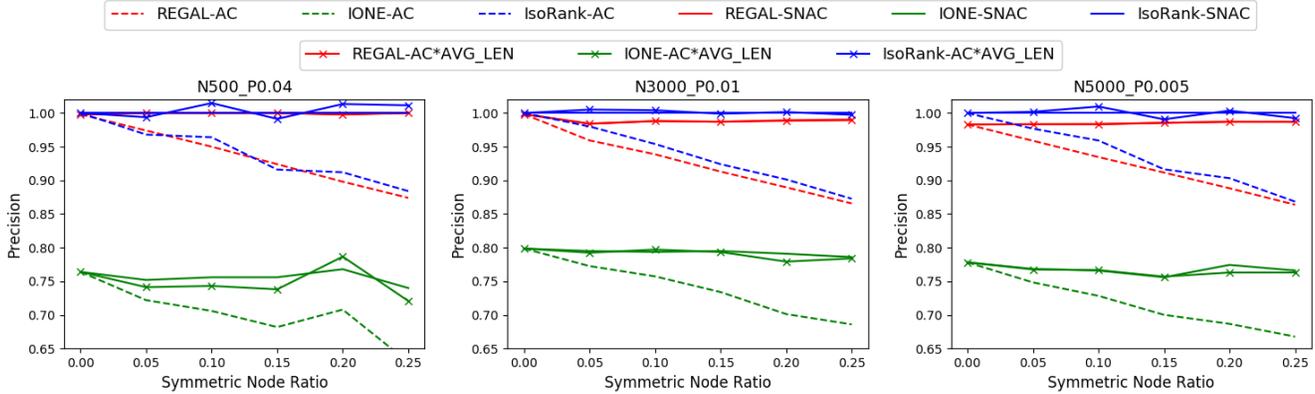

**Fig. 5. Stability Test**

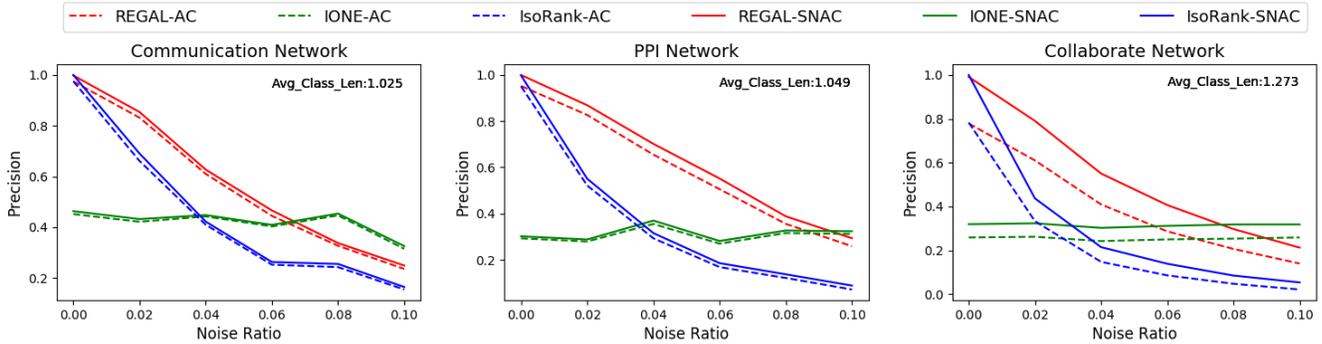

**Fig. 6. Models Performance under SNAC**

### 6.5 Models Performance Under SNAC

In this experiment, we will show the performance of models under SNAC and AC in application scenarios. Following the way of previous literature [2][28] adopted, we randomly remove edges from a given real network with probability $p$, then shuffle the labels of nodes to get the target network.

For cleaner figures, we do not put the additional curve as the stability test does. However, the curve is the same as the stability test.

We can see 2 points from Fig. 6.

Models that do not use anchor links will sharply decline following the increasing noise level, whereas the models leveraging anchor links are minimally affected by the noise, which is demonstrated in previous literature [28].

For the three models, the curve of SNAC is always higher than AC, and the gap is more significant along with the increasing average length of equivalence classes, which is labeled in each subplot of Fig. 6. This shows that the effect of symmetric nodes on AC does exist in practice.

### 6.6 Statistics of Other Datasets

From the analysis in Section 4 and above two experiments, we can see the factor that weakens AC is the average length of equivalence classes $\frac{|V|}{|V/\sim|}$. We collected several common datasets used in network alignment works and compute that length to show the extent of that effect, and the result is listed in Table 3. From the result, we can see that symmetric nodes widely exist in networks.

If not employ SNAC in experiments, researchers can choose datasets with a low level of average length basing on Table1. Researchers can also compute symmetric nodes for data sets other than Table 3 using the algorithm we will publish.

**Table 3. Statistics of Popular Datasets**

| Dataset | Nodes | Edges | Avg Length |
|---|---|---|---|
| Douban-Offline [29] | 1118 | 1511 | 1.336 |
| Douban-Online [29] | 3906 | 8164 | 1.097 |
| Foursquare [7] | 5313 | 54233 | 1.107 |
| Twitter [7] | 5120 | 130575 | 1.02 |
| Elegans [3] | 2995 | 4884 | 1.668 |
| Melanogaster [3] | 7396 | 25054 | 1.157 |
| Sapiens [3] | 10403 | 55168 | 1.118 |
| Arenas [2] | 1135 | 5451 | 1.025 |
| Arxiv [2] | 18772 | 198110 | 1.273 |
| PPI [2] | 3890 | 38739 | 1.05 |
| Dolphins [30] | 62 | 159 | 1.033 |
| Karate Club [31] | 34 | 78 | 1.259 |

## 7. Conclusion

Motivated by the underestimation of models under AC, this paper proposed a novel metric that incorporates identical topology nodes to evaluate network alignment models unbiasedly. We first define such nodes as symmetric nodes. Based on symmetric nodes within single networks and nodes mapping across networks, we give the definition of equivalence classes mapping. The proposed metric SNAC is computed on the consistency between the equivalence classes mapping and the nodes mapping across networks.

We demonstrated SNAC's ability to unbiasedly reflect errors and block the effect of symmetric nodes in both theory and experiments. We further illustrated that the factor determining effect extent is the average length of equivalence classes in datasets and listed that factor of several mainstream datasets to give reference when researchers evaluate models.

Our demonstration is only on undirected, unattributed networks for conciseness, while SNAC can be extended to various settings:

1. It can be directly applied to directed networks without any modification.

2. It can be transplanted to attributed networks. In this case, an equivalence definition of attributes should be given out, we usually adopt *equal* as such equivalence relation.

3. Metric MAP can be combined with SNAC to make it as it does in the information retrieval domain since there can be multiple correct counterparts for each node under SNAC.

**Acknowledgments**

This work was supported by the Fundamental Research Funds for the Central Universities 2020JBM022.

**A: Enhancement from Optimization**

We here show the enhancement of optimization strategies on finding symmetric nodes. We will compare the runtime of the improved method with the way that applies *Bliss* directly on graphs and list the result in Table4. Our experiment is conducted on an AMD Ryzen 5 3500X 6-Core 3.6 GHz system with 16 GB main memory and python2 environment.

The optimization strategies reduce the scale of the problem by dividing the graph into several components, and the nodes distribution of these components dramatically determines the enhancement of optimization. To show the relevance of enhancement to nodes distribution, we put the nodes proportion of the most prominent component reflecting that distribution in Table A. 1.

From Table A.1 we can find that the run time of the finding process dramatically increased as the number of nodes increased; In the case of nodes are primarily clustered in the largest component, the runtime of the improved method is slightly behind the runtime of origin method, this is due to the dividing process. However, for the datasets with a low level of nodes percentage in the largest component, the enhancement of optimization is significant. Mention that the requirements ratio is relatively lenient (for dataset Melanogaster is 98.5%) to reduce the computation.

**Table A. 1 Enhancement from Optimization**

| Datasets | Nodes | Largest Component's Nodes Prop | Original Time(s) | Improved Time (s) |
| --- | --- | --- | --- | --- |
| Arenas [2] | 1135 | 99.8% | 0.108 | 0.203 |
| Arxiv [2] | 18772 | 95.4% | 1307.969 | 777.16 |
| PPI [2] | 3890 | 99.0% | 0.933 | 1.333 |
| Melanogaster [3] | 7396 | 98.5% | 7.168 | 7.01 |
| Sapiens [3] | 10403 | 97.6% | 11.422 | 9.875 |
| Myspace [28] | 10733 | 59.3% | 473.216 | 99.094 |
| Lastfm [28] | 15436 | 86.9% | 1632.112 | 952.019 |

Based on those facts, we can conclude that the optimization strategies significantly improve the computation of the process finding symmetric nodes, especially when the network scale is large and evenly distributed among the components.

**References**

[1] Babai, L., n.d. Groups, Graphs, Algorithms: The Graph Isomorphism Problem 20.
[2] Heimann, M., Shen, H., Safavi, T., Koutra, D., 2018. REGAL: Representation Learning-based Graph Alignment. Proceedings of the 27th ACM International Conference on Information and Knowledge Management 117–126. https://doi.org/10.1145/3269206.3271788
[3] Singh, R., Xu, J., Berger, B., 2008. Global alignment of multiple protein interaction networks with application to functional orthology detection. Proceedings of the National Academy of Sciences 105, 12763–12768. https://doi.org/10.1073/pnas.0806627105
[4] Zhou, F., Liu, L., Zhang, K., Trajcevski, G., Wu, J., Zhong, T., 2018. DeepLink: A Deep Learning Approach for User Identity Linkage, in: IEEE INFOCOM 2018 - IEEE Conference on Computer Communications. Presented at the IEEE INFOCOM 2018 - IEEE Conference on Computer Communications, IEEE, Honolulu, HI, pp. 1313–1321.

https://doi.org/10.1109/INFOCOM.2018.8486231.

[5] Chu, X., Fan, X., Yao, D., Zhu, Z., Huang, J., Bi, J., 2019. Cross-Network Embedding for Multi-Network Alignment, in: The World Wide Web Conference on - WWW '19. Presented at the The World Wide Web Conference, ACM Press, San Francisco, CA, USA, pp. 273–284. https://doi.org/10.1145/3308558.3313499

[6] Shu, K., Wang, S., Tang, J., Zafarani, R., Liu, H., 2017. User Identity Linkage across Online Social Networks: A Review. SIGKDD Explor. Newsl. 18, 5–17. https://doi.org/10.1145/3068777.3068781

[7] Zhang, J., Yu, P.S., n.d. Integrated Anchor and Social Link Predictions across Social Networks 7

[8] Zhang, S., Tong, H., 2016. FINAL: Fast Attributed Network Alignment, in: Proceedings of the 22nd ACM SIGKDD International Conference on Knowledge Discovery and Data Mining. Presented at the KDD '16: The 22nd ACM SIGKDD International Conference on Knowledge Discovery and Data Mining, ACM, San Francisco California USA, pp. 1345–1354. https://doi.org/10.1145/2939672.2939766

[9] Liu, L., Cheung, W.K., Li, X., Liao, L., n.d. Aligning Users Across Social Networks Using Network Embedding 8.

[10] Zhang, S., Tong, H., Maciejewski, R., Eliassi-Rad, T., 2019. Multi-level Network Alignment, in: The World Wide Web Conference on - WWW '19. Presented at the The World Wide Web Conference, ACM Press, San Francisco, CA, USA, pp. 2344–2354. https://doi.org/10.1145/3308558.3313484

[11] Li, C., Wang, S., Wang, Y., Yu, P., Liang, Y., Liu, Y., Li, Z., 2019. Adversarial Learning for Weakly-Supervised Social Network Alignment. AAAI 33, 996–1003. https://doi.org/10.1609/aaai.v33i01.3301996

[12] Ye, R., Li, X., Fang, Y., Zang, H., Wang, M., 2019. A Vectorized Relational Graph Convolutional Network for Multi-Relational Network Alignment, in: Proceedings of the Twenty-Eighth International Joint Conference on Artificial Intelligence. Presented at the Twenty-Eighth International Joint Conference on Artificial Intelligence {IJCAI-19}, International Joint Conferences on Artificial Intelligence Organization, Macao, China, pp. 4135–4141. https://doi.org/10.24963/ijcai.2019/574

[13] Kollias, G., Mohammadi, S., Grama, A., 2012. Network Similarity Decomposition (NSD): A Fast and Scalable Approach to Network Alignment. IEEE Trans. Knowl. Data Eng. 24, 2232–2243. https://doi.org/10.1109/TKDE.2011.174

[14] Emmert-Streib, F., Dehmer, M., Shi, Y., 2016. Fifty years of graph matching, network alignment and network comparison. Information Sciences 346–347, 180–197. https://doi.org/10.1016/j.ins.2016.01.074

[15] Kazemi, E., Hassani, S.H., Grossglauser, M., 2015. Growing a graph matching from a handful of seeds. Proc. VLDB Endow. 8, 1010–1021. https://doi.org/10.14778/2794367.2794371

[16] Man, T., Shen, H., Liu, S., Jin, X., Cheng, X., n.d. Predict Anchor Links across Social Networks via an Embedding Approach 7.

[17] Trung, H.T., Van Vinh, T., Tam, N.T., Yin, H., Weidlich, M., Viet Hung, N.Q., 2020. Adaptive Network Alignment with Unsupervised and Multi-order Convolutional Networks, in: 2020 IEEE 36th International Conference on Data Engineering (ICDE). Presented at the 2020 IEEE 36th International Conference on Data Engineering (ICDE), IEEE, Dallas, TX, USA, pp. 85–96. https://doi.org/10.1109/ICDE48307.2020.00015

[18] Tan, S., Guan, Z., Cai, D., Qin, X., Bu, J., Chen, C., n.d. Mapping Users across Networks by Manifold Alignment on Hypergraph 7.

[19] Garlaschelli, D., Ruzzenenti, F., Basosi, R., 2010. Complex Networks and Symmetry I: A Review. Symmetry 2, 1683–1709. https://doi.org/10.3390/sym2031683

[20] McKay, B.D., Piperno, A., 2013. Practical graph isomorphism, II. arXiv:1301.1493 [cs, math].

[21] Junttila, T., Kaski, P., 2007. Engineering an Efficient Canonical Labeling Tool for Large and Sparse Graphs, in: Applegate, D., Stølting Brodal, G. (Eds.), 2007 Proceedings of the Ninth Workshop on Algorithm Engineering and Experiments (ALENEX). Society for Industrial and Applied Mathematics, Philadelphia, PA, pp. 135–149. https://doi.org/10.1137/1.9781611972870.13

[22] Darga, P.T., Sakallah, K.A., Markov, I.L., 2008. Faster symmetry discovery using sparsity of symmetries, in: Proceedings of the 45th Annual Conference on Design Automation - DAC '08. Presented at the the 45th annual conference, ACM Press, Anaheim, California, p. 149. https://doi.org/10.1145/1391469.1391509

[23] Cordella, L.P., Foggia, P., Sansone, C., Vento, M., 2004. A (sub)graph isomorphism algorithm for matching large graphs. IEEE Trans. Pattern Anal. Machine Intell. 26, 1367–1372. https://doi.org/10.1109/TPAMI.2004.75

[24] Leskovec, J., Kleinberg, J., Faloutsos, C., 2007. Graph evolution: Densification and shrinking diameters. ACM Trans. Knowl. Discov. Data 1, 2. https://doi.org/10.1145/1217299.1217301

[25] Bobby-Joe Breitkreutz, Chris Stark, Teresa Reguly, Lorrie Boucher, Ashton Breitkreutz, Michael Livstone, Rose Oughtred, Daniel H. Lackner, Jürg Bähler, Valerie Wood, Kara Dolinski, Mike Tyers, The BioGRID Interaction Database: 2008 update, Nucleic Acids Research, Volume 36, Issue suppl_1, 1 January 2008, Pages D637–D640, https://doi.org/10.1093/nar/gkm1001

[26] Kunegis, J., 2013. KONECT: the Koblenz network collection, in: Proceedings of the 22nd International Conference on World Wide Web - WWW '13 Companion. Presented at the the 22nd International Conference, ACM Press, Rio de Janeiro, Brazil, pp. 1343–1350. https://doi.org/10.1145/2487788.2488173

[27] Gilbert, E. N., 1959. Random graphs. The Annals of Mathematical Statistics, 30(4), 1141-1144.

[28] Trung, H.T., Toan, N.T., Vinh, T.V., Dat, H.T., Thang, D.C., Hung, N.Q.V., Sattar, A., 2020. A comparative study on

network alignment techniques. Expert Systems with Applications 140, 112883. https://doi.org/10.1016/j.eswa.2019.112883

[29] Zhong, E., Fan, W., Wang, J., Xiao, L., Li, Y., 2012. ComSoc: adaptive transfer of user behaviors over composite social network, in: Proceedings of the 18th ACM SIGKDD International Conference on Knowledge Discovery and Data Mining - KDD '12. Presented at the the 18th ACM SIGKDD international conference, ACM Press, Beijing, China, p. 696. https://doi.org/10.1145/2339530.2339641

[30] Rossi, R.A., Ahmed, N.K., n.d. The Network Data Repository with Interactive Graph Analytics and Visualization 2.

[31] Zachary, W.W., 1977. An Information Flow Model for Conflict and Fission in Small Groups. Journal of Anthropological Research 33, 452–473.